\documentclass{article}

\usepackage{arxiv}

\usepackage[utf8]{inputenc} 
\usepackage[T1]{fontenc}    
\usepackage{hyperref}       
\usepackage{url}            
\usepackage{booktabs}       
\usepackage{amsfonts}       
\usepackage{nicefrac}       
\usepackage{microtype}      
\usepackage{lipsum}

\usepackage{graphicx}
\usepackage{lineno,hyperref}
\usepackage[T1]{fontenc}     
\usepackage{amsmath}
\usepackage{amssymb}
\usepackage{multirow}
\usepackage{subfigure}
\usepackage{graphics}
\usepackage{url}
\usepackage{hyperref}
\usepackage{epstopdf}
\usepackage{longtable}
\usepackage{supertabular,booktabs}

\title{Interactive user interface based on Convolutional Auto-encoders for annotating CT-scans}

\author{
  Martin L\"{a}ngkvist\thanks{Corresponding author} \\
  Center for Applied Autonomous Sensor Systems \\ 
  \"Orebro University \\
  \"Orebro, Sweden
   \And
 Jonas Widell \\
Department of Radiology, Faculty of Medicine and Health \\
\"Orebro University \\
\"Orebro, Sweden
   \And
 Per Thunberg \\
Department of Medical Physics, Faculty of Medicine and Health \\
\"Orebro University \\
\"Orebro, Sweden
    \And
  Amy Loutfi \\
  Center for Applied Autonomous Sensor Systems \\ 
  \"Orebro University \\
  \"Orebro, Sweden
   \And
 Mats Lid\'en \\
Department of Radiology, Faculty of Medicine and Health \\
\"Orebro University \\
\"Orebro, Sweden
}

\begin{document}
\maketitle

\begin{abstract}
High resolution computed tomography (HRCT) is the most important imaging modality for interstitial lung diseases, where the radiologists are interested in identifying certain patterns, and their volumetric and regional distribution. The use of machine learning can assist the radiologists with both these tasks by performing semantic segmentation. In this paper, we propose an interactive annotation-tool for semantic segmentation that assists the radiologist in labeling CT scans. The annotation tool is evaluated by six radiologists and radiology residents classifying healthy lung and reticular pattern i HRCT images. The usability of the system is evaluated with a System Usability Score (SUS) and interaction information from the readers that used the tool for annotating the CT volumes. It was discovered that the experienced usability and how the users interactied with the system differed between the users. A higher SUS-score was given by users that prioritized learning speed over model accuracy and spent less time with manual labeling and instead utilized the suggestions provided by the GUI. An analysis of the annotation variations between the readers show substantial agreement (Cohen's kappa=0.69) for classification of healthy and affected lung parenchyma in pulmonary fibrosis. The inter-reader variation is a challenge for the definition of ground truth.
\end{abstract}


\section{Introduction}
The application of machine learning in radiology has gained a lot of interest in recent years in various fields of medical image analysis~\cite{LITJENS201760}. In computed tomography imaging (CT), a rotating x-ray tube and detector are used to acquire thin slice images of different parts of the body. In a particular application, high resolution CT of the lungs (HRCT), has during the last decades emerged as the most important image modality for interstitial lung diseases, including pulmonary fibrosis~\cite{Hansell2008,Raghu2011}.

The radiologists analyze the HRCT images through visual assessment, by identifying the presence and distribution of pathological patterns in the lung parenchyma. The radiologists are therefore both interested in identifying certain patterns (classification), and assessing the distribution of the patterns (segmentation).

Semantic segmentation is the task of classifying all pixels in an image and can solve both the classification and the segmentation task. Various image-decoders networks have recently shown to be successful as image-decoders for solving the semantic segmentation task, such as variational auto-encoder~\cite{2013arXiv1312} and generative adversarial network (GAN)~\cite{NIPS2014_5423}.

However, these models have a high number of parameters and require a large amount of labeled data. Acquiring labeled HRCT data is costly, time-consuming, and requires the expertise from radiologists. In this study, we instead present an interactive graphical user-interface (GUI) including a convolutional auto-encoder that aims to assist the radiologist in the process of annotating CT images in order to speed-up the labeling process.

The inter-reader variation in the interpretation of HRCT is a well known issue in the visual analysis  ~\cite{Walsh2016,Watadani2013}. This is also the rational for the development of quantitative tools for HRCT analysis ~\cite{Bartholmai2013,Jacob2016,Humphries2017}. Quantitative analysis of pulmonary fibrosis is based on the delineation of affected and healthy lung parenchyma, which requires a definable border between healthy and affected parenchyma to be present. However, to the best of our knowledge, there is no previous study that quantifies the inter-reader variability in the delineation of healthy and affected lung parenchyma in HRCT images.

The first purpose of the present study is, therefore, to evaluate the usefulness of the interactive GUI for the labeling process and the second purpose is to quantify the reader variations in delineation of healthy lung and reticular pattern in HRCT using the developed GUI.

\section{Related Work}
\label{sec:related}

There has been a large amount of effort into developing tools for annotating images~\cite{yuen2009labelme} and videos~\cite{mihalcik2003design,ambardekar2009ground} for computer vision applications and research. Many different strategies have been explored in order the reduce the human manual effort and required annotation time. The work by~\cite{all2011flowboost} proposes doing sparse labeling of a video and the tool suggests labels for the inbetween frames based on the current detector and physical constraints on target motions.

Semi-automatic tools that provide hotkey-shortcuts, smart drawing tools, drag-and-drop, and integrated computer vision segmentation algorithms allow humans to annotate more efficiently~\cite{kavasidis2012semi}. The use of semi-supervised learning (see~\cite{zhu2006semi} for an overview) makes use of the large amount of available unlabelled data and has been used for object recognition~\cite{fergus2009semi,lee2011learning} and in an active learning setting~\cite{zhou2006enhancing,joshi2010breaking}.

The use of transfer learning~\cite{pan2010survey,lampert2009learning} allows pre-trained models to share knowledge and provide suggestions during the human supervised annotation of new data. The work by~\cite{viraj2018} uses transfer learning with a pre-trained RetinaNet model for an object detection annotation tool. 

The authors in~\cite{vijayanarasimhan2011cost} and~\cite{vijayanarasimhan2014large} have developed methods that effectively rank the query images and prioritize images that gives the best tradeoff between human effort and the information gain in order to reduce the needed human intervention. 

An alternative to using AI tool-assisted programs is to use crowd-sourcing annotation~\cite{sorokin2008utility,welinder2010online} and has been used for videos~\cite{vondrick2013efficiently,kavasidis2014innovative}, object detection ~\cite{kavasidis2013generation}, and energy data~\cite{7364072} with the use of web-based frameworks or online games. 

In this work, we focus on an interactive, semi-automatic annotation tool for semantic segmentation of CT-scans. Semi-automatic annotation has previously been explored an other datasets such as cooking videos~\cite{semicookingvideo} and soccer videos~\cite{5280004}. 

Another annotation tool for object detection in videos that has been proposed is the interactive Video Annotation Tool, iVAT~\cite{BIANCO201588} that supports manual, semi-automatic, and automatic annotations. The tool can train a model from a list of computer vision algorithm on already labeled data in an incremental learning framework. The work uses a quantitative evaluation and system usability study. Similarly, the work by~\cite{vondrick2011video} uses active learning to query the user for corrections and shows that the amount of human effort, measured in the number of user interactions, is reduced. 

While these works uses active learning and semi-automatic setups, they are not strictly interactive since the user needs to wait for a complete training pass before getting feedback. An interactive object annotation method has been proposed by~\cite{yao2012interactive} but is more focused on incrementally train the object detector while the user provides annotations in order to reduce human annotation time rather than learning performance.

There are many interactive segmentation methods for images that have been proposed~\cite{boykov2001interactive,criminisi2008geos,rajchl2017deepcut}, and specifically for medical image segmentation ~\cite{zhao2013overview,wang2018interactive}

Our work differs from these previous works in a number of ways. We focus on a fully semantic segmentation of the data and not on annotating bounding boxes for object detection or segmentation in medical images. There is also a larger focus on evaluating the experienced system usability and user interaction and relate the inter- and intrareader variation to the perceived usability of the system. The focus on interactivity and continuously provided feedback allows us to examine the difference between how users used the system and their preferred trade-off between training speed and accuracy in the provided feedback.

\section{Materials and methods}

\subsection{Interactive GUI for annotation}
\label{sec:interactiveGUI}

\begin{figure}[!ht]
\centering
\includegraphics[width=0.8\textwidth]{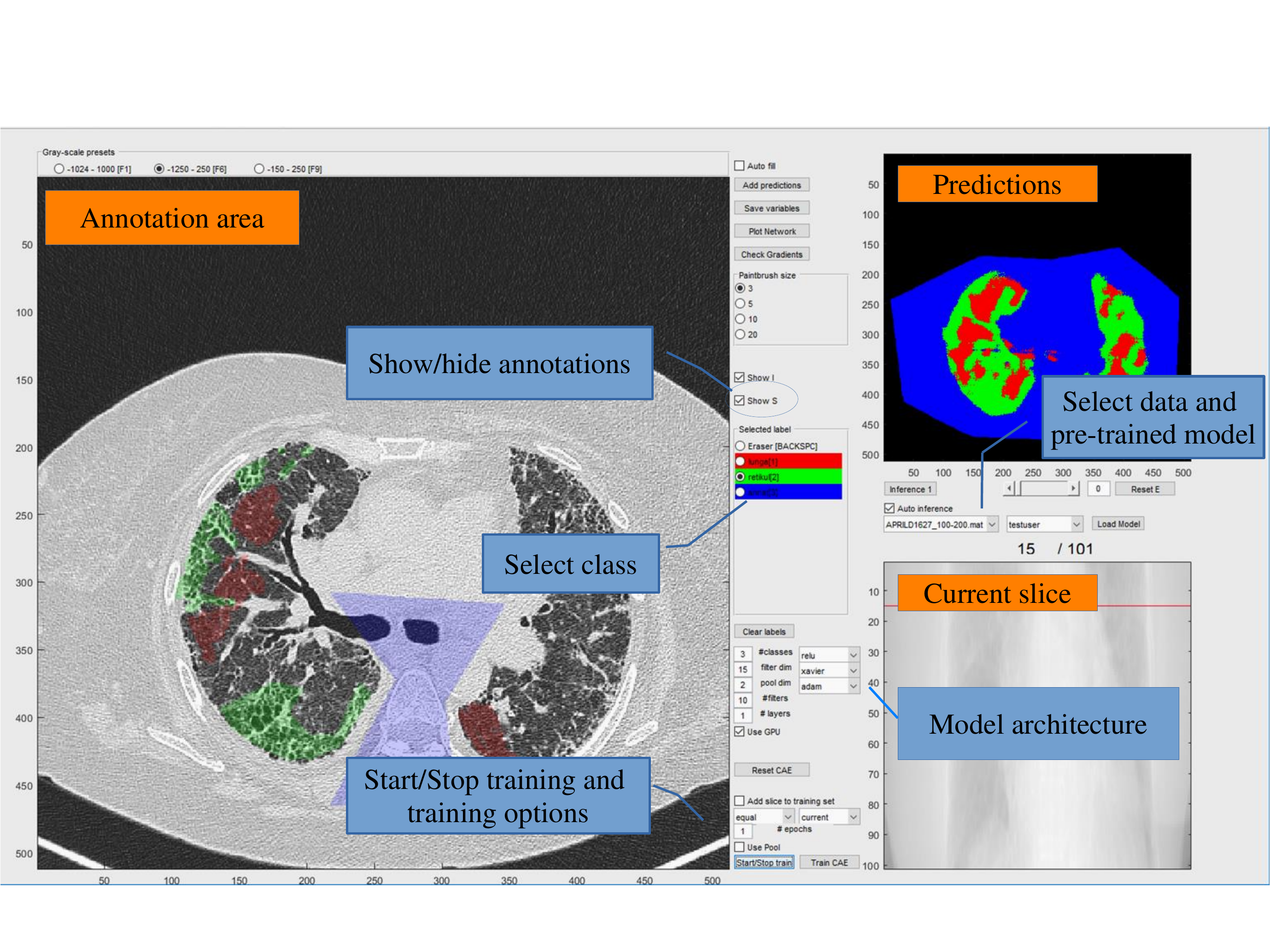}
\caption{GUI used for annotating and training the model. The CT data and the labels are shown in the \emph{annotation area}. The user labels data by selecting class and brush size and marks the data. The predictions from the model is shown in the \emph{prediction area}. If the user is satisfied with the output from the model, the predictions can be added to the \emph{annotation area} and later make corrections where necessary. The user can change slices by clicking the \emph{current slice area} or using the mouse wheel. The user can define the models hyperparameters, such as number of classes, filter dimensions, pooling size, number of layers, and number of filters in each layer in the \emph{model architecture area}. Showing annotations and/or raw data can be toggled. The predictions under a threshod controlled by a draggable slider can be hidden.}
\label{fig:guioversikt}
\end{figure}

An interactive graphical user-interface (GUI) was developed for this study, see Figure~\ref{fig:guioversikt}. The GUI works as an annotation tool where the user can annotate CT volumes effectively and is designed with a radiologist's feedback to resemble the look and functionality of traditional software used by radiologists. In addition, the GUI contains a fast classifier that is concurrently training in the background to make predictions during the labeling process, which gradually improves as more pixels become labeled. The GUI also stores information about how the user is using the interface for evaluation purposes. The user can draw with different brush sizes in the 2D scans or fill polygons in 2D and 3D. The learning process is iterative and the model is reset for each new user so that each user gets a custom-trained model based on only their own annotations. 

\subsection{Convolutional Auto-encoder}
\label{sec:cae}

\begin{figure}[!ht]
\centering
\includegraphics[width=0.95\textwidth]{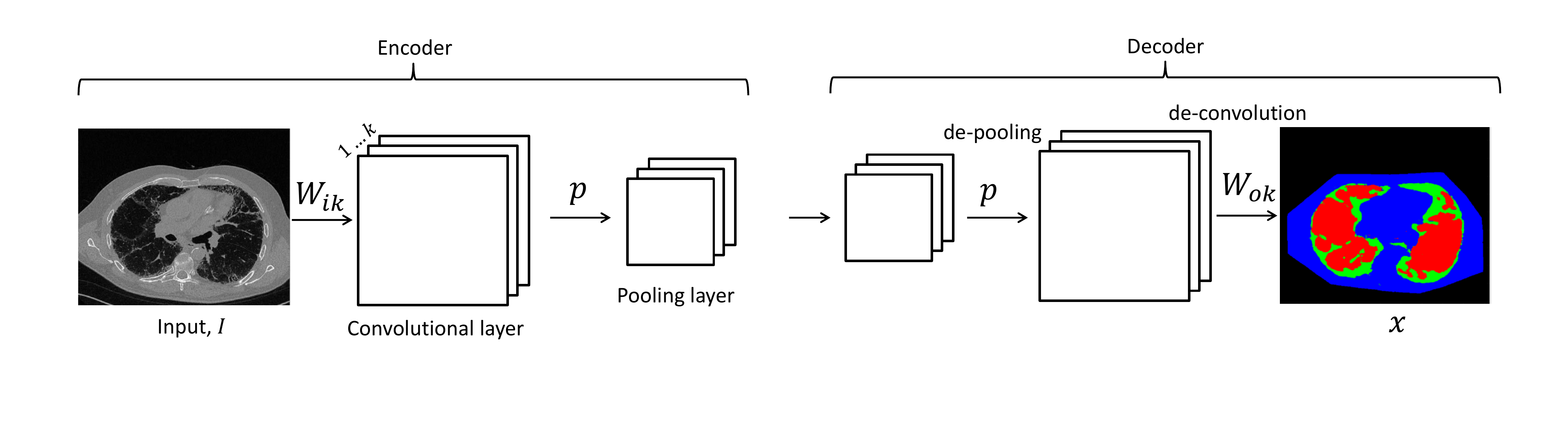}
\caption{Overview of a one layer of Convolutional Autoencoder (CAE) that consists of an encoder and a decoder. The input is slice of a CT scan and the output is an image of the classified pixels.}
\label{fig:CAEoverview}
\end{figure}

A Convolutional Auto-encoder (CAE)~\cite{masci2011stacked} is used to perform the classification of the CT images and consists of a encoder and a decoder, see Figure~\ref{fig:CAEoverview}. The encoder calculates the k-th feature map in the convolutional layer as:
\begin{equation}
h^k = \sigma\left(I \ast W_1^k + b^k\right)
\end{equation}
where $I$ is the gray-scale input image, $W_1^k$ is the k-th filter between the input and convolutional layer, $b^k$ is the bias for the k-th filter, $\ast$ is the convolution operation, and $\sigma$ is the Rectified Linear Unit (ReLu)~\cite{nair2010rectified} activation function. Downsampling of the convolutional layer is then performed by taking the maximum value in each $p\times p$ non-overlapping subregion. 

The decoder transforms the representation from the encoder back to the original size by first unpooling the representation and performs a deconvolutional operation as:
\begin{equation}
\hat{y} = softmax\left(h^k \ast W_2^k + c^K\right)
\end{equation}
where $W_2^k$ is the k-th filter from the unpooling layer the output $\hat{y}$, $c^K$ is the bias for the K-th output layer. 

The architecture for the CAE used in this work was a one-layered CAE with 10 filters, filter size 15, and pool dimension 2. The design choices are motivated by the aim of achieving a fast classifier for interactive use. 

\subsection{Image Data}
\label{sec:imagedata}
Twelve scans, demonstrating pulmonary fibrosis with reticular pattern and honeycombing, were selected from a retrospectively created databank consisting of high resolution CT of the lungs (HRCT). The images were acquired with Siemens Somatom Definition AS (n=6), Siemens Somatom Definition Flash (n=4) and Siemens Biograph (n=2). Images were reconstructed as contiguous 1 mm slices using a high spatial resolution kernel (B70f or I70f1). Aquisition parameters were 120 kVp (n=9), 100 kVp (n=2), 140 kVp (n=1), and standard chest reference mAs settings (ref-mAs 150-160, ref-kV 120).

\subsection{Image Review}
\label{sec:imagereview}

Two radiologists (seven and five years of experience in thoracic radiology) and four radiology residents participated in the study. The study was performed with two separate experiments. In the first reading, where all readers participated, the user interface in addition to the inter- and intra-reader variation, was evaluated. The purpose of the second reading, where only the two thoracic radiologists participated, was to obtain additional data for the inter-reader variation.

The task in the experiments was to completely annotate four specific slices, interspaced with 1 cm, through the middle part of the lungs in the three categories normal lung parenchyma; reticular pattern including honeycombing; and non-pulmonary tissue. The same image slices were annotated by all readers. Structures and air outside the body were not annotated. 

Immediately before the first reading, the readers conducted a tutorial designed to guide the reader through the central and necessary functions of the interactive GUI. One reader (reader 6) only conducted a written, unsupervised, tutorial and one reader (reader 3) was already familiar with the GUI. The other readers conducted a supervised written tutorial, that was completed in approximately 30 minutes.

In the first reading, each reader annotated four slices in four CT stacks. The first stack was repeated as number four for the analysis of intra-reader variation. Some readers completed the annotation during a single session, while others completed the annotation over multiple sessions. After completing the annotation task, each reader received a standardized questionnaire, System Usability Score (SUS).

In the second reading, the two thoracic radiologists performed an additional annotation task on nine more CT-stacks. The task in the second reading was identical to the first reading. Since the purpose of the second reading was to obtain inter-reader variation data on a larger number of subjects, the usability of the system was not assessed.

\subsection{Inter- and intrareader variability analysis}
\label{sec:interreader}

The inter- and intrareader variations in the annotation between the six readers in the first reading were evaluated on a pixelwise basis, and were quantified with Cohen's kappa and Jaccard index (Intersection over Union), that were computed for each pair of readers, using all slices in the first three CT stacks. Cohen's kappa was computed for the three-class classification healthy praenchyma, reticular pattern or non-pulmonary tissue, and for the two-class classification healthy parenchyma or reticular pattern.

The intra-reader analysis was performed for each reader using the annotations performed on the identical first and fourth CT stacks. Pixelwise Cohen's kappa and Jaccard index were computed similar to the inter-reader analysis for each reader with 95\% confidence interval.

For the two readers that performed the second annotation task, the inter-reader variations were correspondingly quantified using all twelve CT stacks.

The Regional Research Ethics Board approved the study protocol and waived the informed consent requirement for images in the data bank.


\section{Experimental results}

\subsection{GUI evaluation}
\label{sec:res_GUI}

The usability of the GUI was evaluated using the System Usability Scale (SUS)~\cite{Brooke1996}. The survey consists of 10 statements that uses a Likert scale where the answers are graded from 1 (strongly disagree) to 5 (strongly agree). The combination of the answers measures the user's effectiveness, efficiency, and satisfaction and a final score between 0 and 100 is calculated and converted to letter grades for easier interpretation~\cite{bangor2009determining}.

Table~\ref{table:susscore} shows the survey results from the six radiologists used in this study. A SUS-score above 68 (grade D) is considered above average. Half of the users received a SUS-score below the usability threshold and the other half received a higher value, meaning that there is a disagreement between the radiologist regarding the usability of the system. 

\begin{table}[!ht]
\caption{Calculated SUS-scores from six readers and their respective grades (A-F).}
\label{table:susscore}
\centering
\begin{tabular}{ccc}
Reader & SUS-score & Grade\\ \hline
\#1 & 58 & F \\ 
\#2 & 75 & C \\ 
\#3 & 73 & C \\ 
\#4 & 53 & F \\ 
\#5 & 68 & D \\ 
\#6 & 40 & F\\ \hline 
Average & $60.8 \pm 13.4$
\end{tabular}
\end{table}

\begin{figure}[!ht]
\centering
\includegraphics[width=0.95\textwidth]{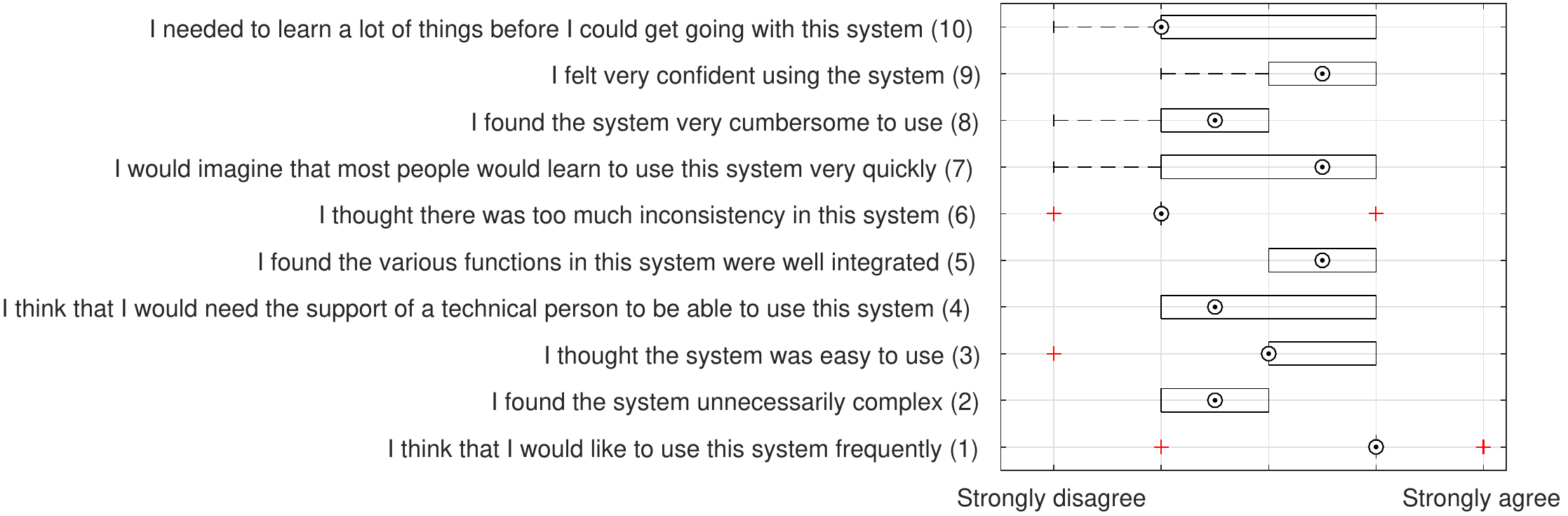}
\caption{A box plot of the answers from six radiologists to the 10 questions in the SUS survey. The boxes are the $25th$ and $75th$ percentiles, dashed lines are the min and max values, circle indicate median, and the plus symbol indicate outlier.}
\label{fig:susboxplot}
\end{figure}

The results can further be analysed by examining the answers from the individual questions in the SUS survey. Figure~\ref{fig:susboxplot} shows a box plot of the answers to each question in the SUS-survey. 

Most users answered that they would use this system frequently (1), except one user that thought that the system was not easy to use (3). The other users thought the system was somewhere between easy and difficult to use (2), (4), and (8). Most users agree that the functions in the system were well integrated and the users felt very confident using the system (5), (9).


\subsection{User interaction evaluation}

\begin{figure}[!ht]
 \centering
    \subfigure[]{\includegraphics[width=0.31\textwidth]{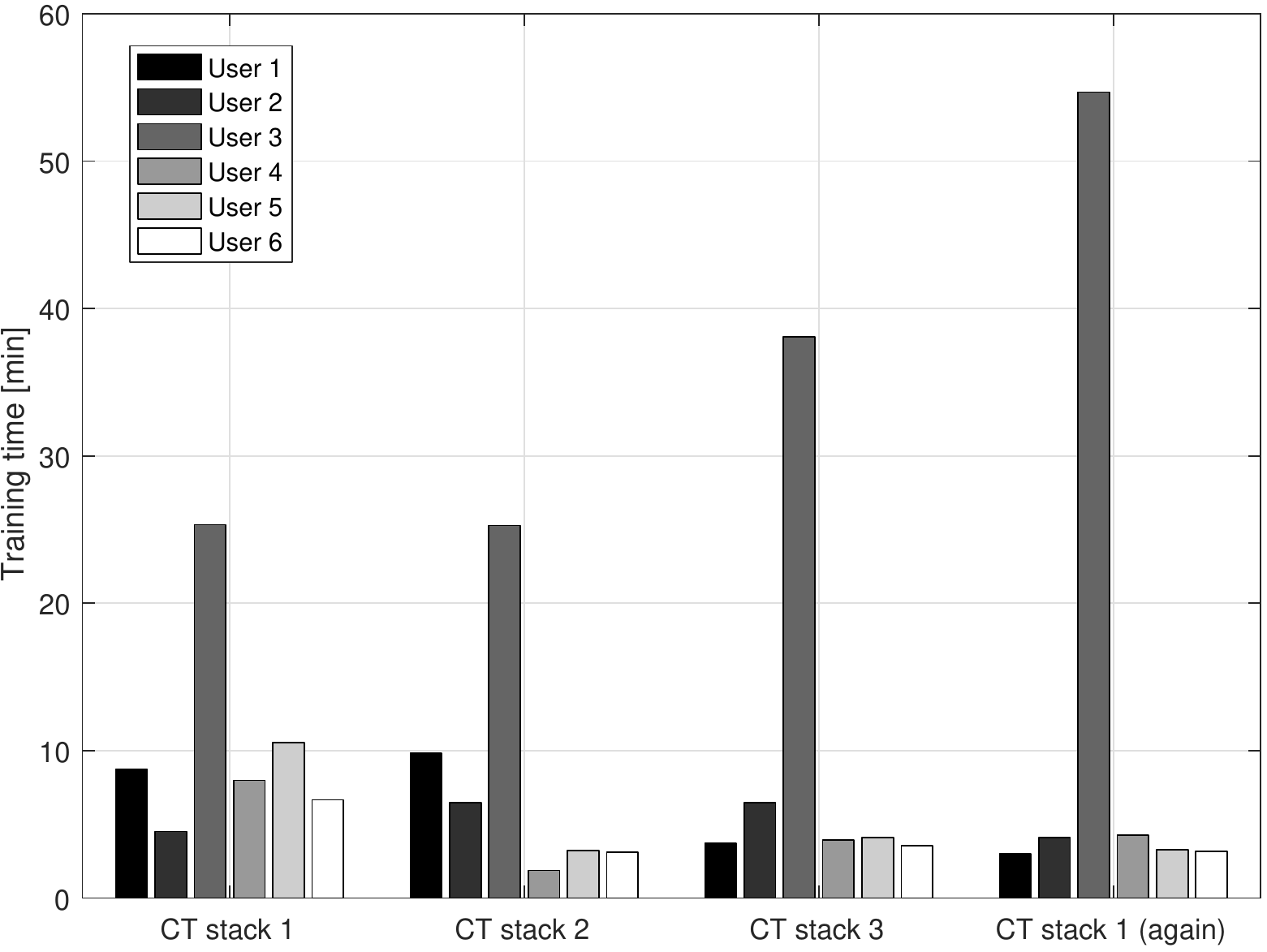} \label{fig:traintime}}
        \subfigure[]{\includegraphics[width=0.31\textwidth]{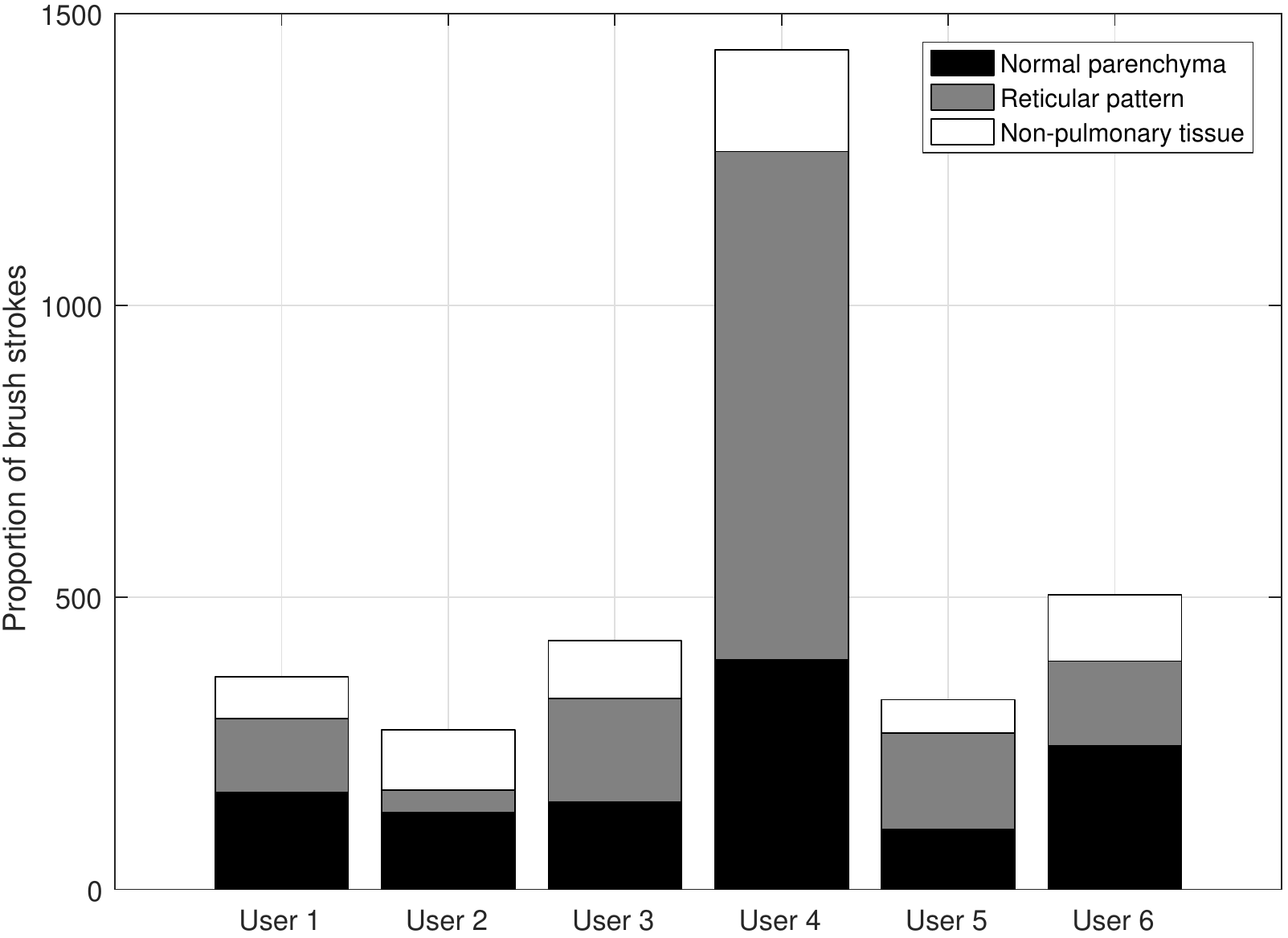} \label{fig:proportion}}
    \subfigure[]{\includegraphics[width=0.31\textwidth]{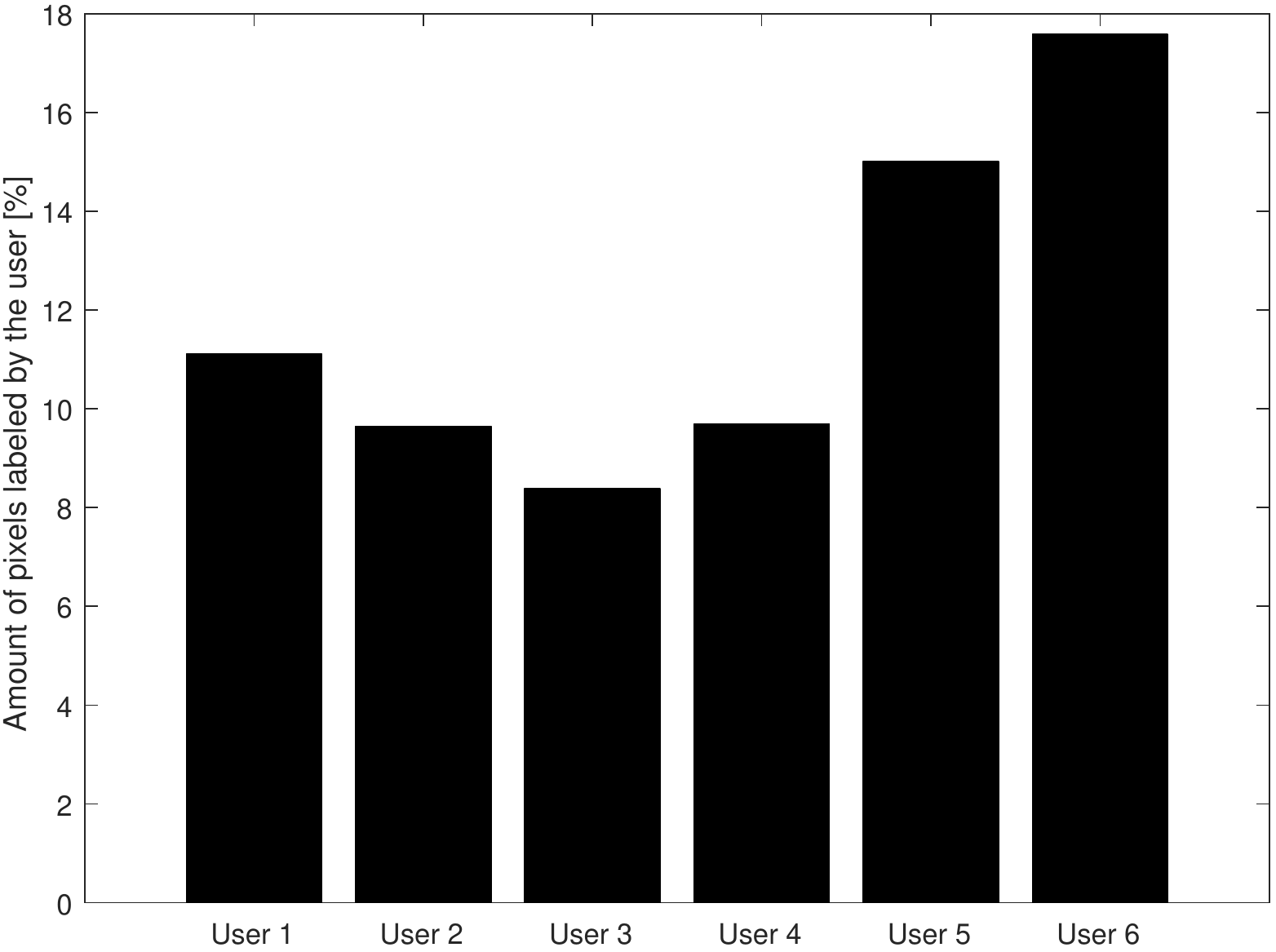} \label{fig:perc_stack1}}
    \caption{(a). Training time for each user for each stack. (b) Amount of manual labeling in terms of number of brush strokes for each user colored by the class for the first stack. (c) Percentage of the data that was labeled by the user for the first stack.}
  \label{fig:interaction}
\end{figure}

The system logged the total labeling time, percentage of data labeled by the user, amount of manual work in terms of initial labeling and corrections, see Figure~\ref{fig:interaction}. All users labeled the CT stacks in the same order (CT stack 1, 2, 3, and then 1 again). The model started from random initialization and was only reset when a new user started using the GUI. This resulted in each user getting a personalized trained model that was only trained on that users annotated data. 

Figure~\ref{fig:traintime} shows how long the model was training for each user for each stack. The training time is the highest for the first stack and decreasing for each stack, except for user 3 due to interruptions during labeling. The average training time for all users, except user 3, was for the first and last stack $7.7$ minutes and $3.6$ minutes, respectively.

Figure~\ref{fig:proportion} shows the amount of brush strokes for each user on the first stack colored by class. User 4 used the most brush strokes compared to the other users. The average proportional amount of brush strokes over all users and all stacks are $36.4\pm7.7\%$, $38.9\pm13.4\%$, $24.8\pm8.9\%$, for normal parenchyma, reticular pattern, and non-pulmonary tissue, respectively. Meaning that most manual work is spent on labeling normal parenchyma and reticular pattern, which are the most difficult classes to differentiate.

Figure~\ref{fig:perc_stack1} shows how large percentage of the first stack was labeled by the user. All users labeled on average $12\%$ of the data for the first stack and the GUI labeled the rest. User 3 labeled the least amount with $8\%$ and user 6 the most with $18\%$.


\subsection{Inter- and intrareader variation}
\label{sec:res_interreader}

Quantitative measures for the inter- and intrareader variations are shown in Table~\ref{table:interreadertable1} and Table~\ref{table:interreadertable2}. As anticipated, the agreement on the delineation of non-pulmonary tissue is excellent with a Jaccard index of close to 1. The variation in labels between the readers was consequently related to the delineation between healthy and affected lung parenchyma.

Cohen's kappa analysis showed excellent agreement (kappa=0.86) between readers for the three-class analysis. When the non-pulmonary tissue, which is diagnostically unproblematic, was removed from the analysis, the agreement was substantial (kappa = 0.69).

\begin{table}[!ht]
\centering
\caption{Inter- and intra-reader variation, Jaccard index}
\label{table:interreadertable1}
\begin{tabular}{lccc}
\hline
& \textbf{\begin{tabular}[c]{@{}c@{}}Normal\\ parenchyma\end{tabular}}        
& \textbf{\begin{tabular}[c]{@{}c@{}}Reticular\\ pattern\end{tabular}}       
& \textbf{\begin{tabular}[c]{@{}c@{}}Non-pulmonary\\ tissue\end{tabular}}    \\ 
\hline

\begin{tabular}[c]{@{}l@{}}\textbf{Inter-reader variation}\\ \textit{(6 readers, 3 CTs)}\\mean (95\%CI){[}Range{]}\end{tabular}
& \begin{tabular}[c]{@{}c@{}}0.81 (0.79-0.82)\\ {[}0.75-0.84{]}\end{tabular} 
& \begin{tabular}[c]{@{}c@{}}0.58 (0.55-0.60)\\ {[}0.49-0.65{]}\end{tabular} 
& \begin{tabular}[c]{@{}c@{}}0.94 (0.92-0.95)\\ {[}0.89-0.97{]}\end{tabular} \\

& & & \\

\begin{tabular}[c]{@{}l@{}}\textbf{Inter-reader variation}\\\textit{(2 readers, 12 CTs)}\end{tabular}
& \begin{tabular}[c]{@{}c@{}}0.76\end{tabular} 
& \begin{tabular}[c]{@{}c@{}}0.58\end{tabular} 
& \begin{tabular}[c]{@{}c@{}}0.97\end{tabular} \\

& & & \\

\begin{tabular}[c]{@{}l@{}}\textbf{Intra-reader variation}\\ \textit{(6 readers, 1 CT)}\\mean (95\%CI){[}Range{]}\end{tabular}
& \begin{tabular}[c]{@{}c@{}}0.82 (0.76-0.87)\\ {[}0.77-0.88{]}\end{tabular}
& \begin{tabular}[c]{@{}c@{}}0.62 (0.47-0.68)\\ {[}0.47-0.68{]}\end{tabular}
& \begin{tabular}[c]{@{}c@{}}0.94 (0.87-1.00)\\ {[}0.81-0.97{]}\end{tabular}          \\ 
\hline
\end{tabular}
\end{table} 

\begin{table}[!ht]
\centering
\caption{Intra- and inter-reader variations, Cohen's kappa}
\label{table:interreadertable2}
\begin{tabular}{lcc}
\hline

& \textbf{\begin{tabular}[c]{@{}c@{}}Three-class\\ Cohen's kappa\end{tabular}}
& \textbf{\begin{tabular}[c]{@{}c@{}}Two-class\\ Cohen's kappa\end{tabular}}          \\ 
\hline

\begin{tabular}[c]{@{}l@{}}\textbf{Inter-reader variation}\\\textit{(6 readers, 3 CTs)}\\mean (95\%CI){[}Range{]}\end{tabular} & \begin{tabular}[c]{@{}c@{}}0.86 (0.84-0.87)\\ {[}0.81-0.88{]}\end{tabular} 
& \begin{tabular}[c]{@{}c@{}}0.69 (0.67-0.71)\\ {[}0.61-0.75{]}\end{tabular} \\

& & \\

\begin{tabular}[c]{@{}l@{}}\textbf{Inter-reader variation}\\\textit{(2 readers, 12 CTs)}\end{tabular} & \begin{tabular}[c]{@{}c@{}}0.85\end{tabular} 
& \begin{tabular}[c]{@{}c@{}}0.65\end{tabular} \\

& & \\

\begin{tabular}[c]{@{}l@{}}\textbf{Intra-reader variation}\\ \textit{(6 readers, 1 CT)}\\mean (95\%CI){[}Range{]}\end{tabular} & \begin{tabular}[c]{@{}c@{}}0.87 (0.83-0.90)\\ {[}0.80-0.90{]}\end{tabular}
& \begin{tabular}[c]{@{}c@{}}0.72 (0.65-0.79)\\ {[}0.62-0.80{]}\end{tabular}          \\ 
\hline

\end{tabular}
\end{table} 

In the second reading, where two readers provided labels on twelve CTs, the reader variations were similar as in the first reading with excellent agreement in the three-class analysis and substantial agreement in the two-class analysis, see Table~\ref{table:interreadertable1} and Table~\ref{table:interreadertable2}. Also, the analysis of intra-reader variations showed similar Jaccard index and Cohen's kappa as in the inter-reader analysis. However, the numbers are not entirely comparable, since the intra-reader variation was analyzed on just one CT stack, while the inter-reader variations were computed using annotation data from three or twelve different CT stacks.

The inter-reader variations in the delineation of healthy and reticular lung parenchyma can to a large part be explained by the personal cut-off for each reader on the continuous spectrum between clearly healthy lung parenchyma and typical reticular pattern. 

\begin{figure}[!ht]
\centering
\includegraphics[width=0.95\textwidth]{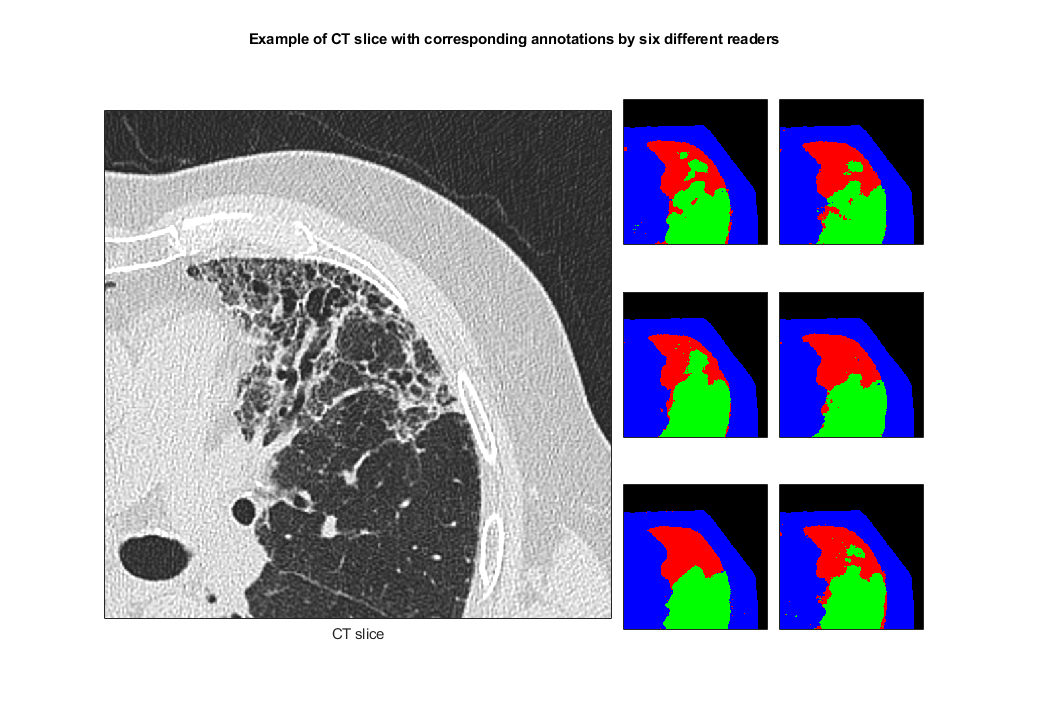}
\caption{Detail from one annotated CT slice with corresponding labelling by six differnt readers (Red - reticular pattern; Green - normal parechyma; Blue - non-pulmonary tissue). }
\label{fig:interrater}
\end{figure}

Figure \ref{fig:interrater} shows an example on one of the annotated slices of the delineation by six different readers.

\subsection{Model classification accuracy}

The capacity of the model determines the trade-off between accuracy and training time. The model should aim to reduce the training time by using a small model in favor of classification accuracy during the labeling process. Once enough data has been labeled, a larger model can be trained to achieve higher performance. In this section we compare the classification accuracy between different size models. The training data are the slices 5, 15, and 25 for all of the three stacks. The test data is slice 35 for all three stacks. The ground-truth is taken as the mode of the annotation from all 6 readers. Table~\ref{table:classacc} shows how the total accuracy is increasing with larger model sizes. The F1-score for each class is also shown and is increasing for all classes with increased number of layers. The most difficult class to classify is reticular pattern, followed by normal praenchyma. The easiest class to classify is non-pulmonary tissue. The F1-score is measurement of the balance between precision and recall and is calculated as:
\begin{equation}
F_1 = 2 \times \frac{precision*recall}{precision+recall}.    
\end{equation}


\begin{table}[!ht]
\caption{Classification accuracy, F1-score per class (normal praenchyma, reticular pattern and non-pulmonary tissue), and training time for different number of layers in the classifier.}
\label{table:classacc}
\centering
\begin{tabular}{cccc}
$\#$layers & Accuracy [$\%$] & F1-score & Training time [h] \\ \hline
1 & 88.4 & 0.79, 0.69, 0.91 & 1 \\
3 & 91.4 & 0.85, 0.75, 0.93 & 2 \\
5 & 93.3 & 0.87, 0.83, 0.96 & 7 \\
\end{tabular}
\end{table}




    
\section{Discussion}
Acquisition of labeled image data is a key issue for the development of machine learning methods in medical imaging. The aquisition process is costly and time consuming, especially in medical imaging, where labels need to be provided by medical expert readers~\cite{Choy2018}. In the present study, we propose an interactive GUI using a convolutional autoencoder for pixelwise labelling of CT volumes, and we use the GUI to compute the inter- and intra-reader variations in the delineation of pathology.

The definition of ground truth is crucial for the results of the neural network training, and in semantic segmentation the labels need to be provided pixelwise. In the present study, the delineation between healthy and affected (reticular) lung parenchyma in HRCT in patients with pulmonary fibrosis was studied~\cite{Hansell2008}. The analysis of inter-reader variation showed that the different readers had different visual cut-off between affected and healthy lung parenchyma, see Figure \ref{fig:interrater}. The inter-reader variation in the interpretaion of medical images including HRCT in idiopathic pulmonary fibrosis has been analyzed before~\cite{Walsh2016,Watadani2013}. While previous studies focus on a more general view, the present study demonstrates the inter-reader variation in the detailed delineation of healthy and affected parechyma in the images.

\begin{figure}[!ht]
\centering
\includegraphics[width=0.95\textwidth]{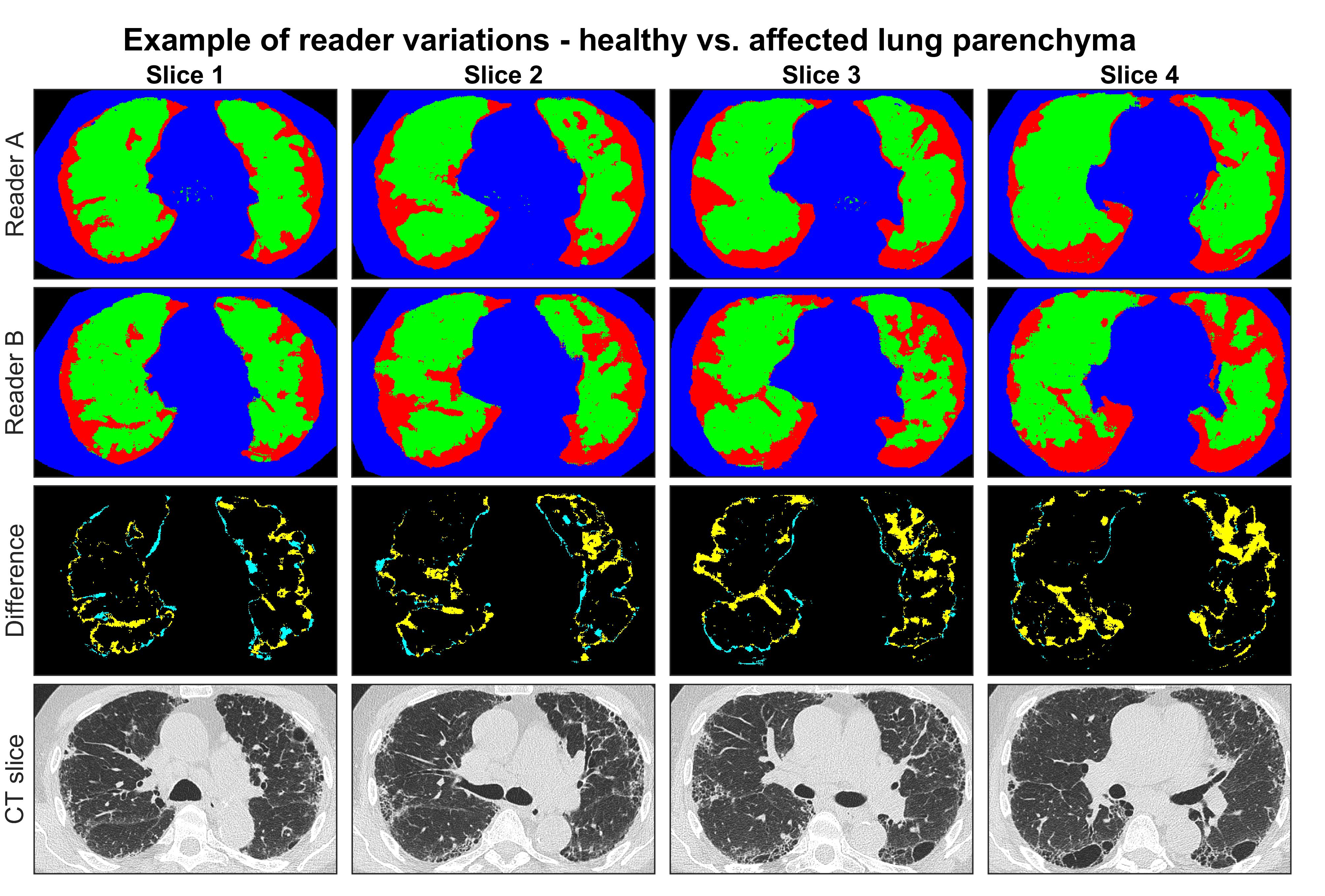}
\caption{Example of reader variations in one CT stack in the second reading. Green, red, blue - labelled as healthy lung, affected lung and non-pulmonary tissue. Yellow - Labelled normal by reader A and affected by reader B. Cyan - Labelled affected by reader A and normal by reader B.}
\label{fig:interrater2}
\end{figure}

The inter-reader variations were quantified in two experimental designs - between several readers on a small number of CT stacks and between two readers on a larger number of examinations. As detailed in Table~\ref{table:interreadertable1} and Table~\ref{table:interreadertable2}, the quantified inter-reader variation was consistent in the two experiments. A visual analysis of the labels provided by the two radiologists in the second reading, shows that the spatial distribution of the provided labels are similar between the readers, see Figure \ref{fig:interrater2}. The difference in the provided labels, quantified in the inter-reader analysis, is essentially the position of the boundary, but not the location of the of the affected areas.

The different cut-offs between the readers for delineating affected parenchyma in pulmonary fibrosis suggests that  there is no discrete border between healthy and affected parenchyma, but rather a continuous range between healthy and clearly pathologic lung pattern. This finding has important points concerning volumetric analysis of pathology in HRCT: First, an objective reader independent method is necessary for any study involving volumetric analysis of HRCT. Second, a thorough validation of any reader independent method is necessary to ensure consistency. Third, a perfect reader independent method is not reasonably achievable.

A strength in the study is the inclusion of several radiologists and radiology residents that provided detailed annotations in the included slices. Detailed annotation of scattered findings such as reticular parenchyma, as demonstrated in Figure \ref{fig:interrater2}, is time-consuming, and to preserve the detailed annotation we intentionally limited the number of slices per examination. 

In the interactive labelling phase, learning speed was preferred over accuracy which motivated the use of a single layered network. The development of more complex models, after acquisition of labeled data, verified that when training time is not crucial, a better performing system can be achieved. The study thus demonstrated how the machine learning algorithms used in the labelling, and those used for final model development can be separated depending on their purpose.

The way that the users interacted with the system, and the perceived usability of the system, differed from the different users. Users that spent less time with manual labeling and utilized the predictions from the classifier reported a higher usability SUS-score. For future work, the system needs to be simplified to make it easier to use and more optimized for performance and faster training.

There are some limitations in the present study. The readers that were included from a single radiology department, only received a short training before performing the test. With more acquaintance, the interaction with the GUI may change. Although the delineation task and annotation classes~\cite{Hansell2008} were clearly defined for the readers, the design of the GUI may, just like any other tool, influence the labels provided by readers, which can affect the measured inter-reader variability.

\section{Conclusion}
The present study demonstrated how a fast convolutional auto-encoder can be used interactively when pixel-wise labels of CT-scans are acquired, while a more complex model can be used when training time is not an issue. The inter-reader variability is an obstacle for the definition of ground truth, but also motivates the development of AI tools that may improve quantitative image analysis in HRCT.

\section*{Acknowledgements}
This work has been sponsored by Nyckelfonden (grant OLL-597511) and by Vinnova under the project Interactive Deep Learning for 3D image analysis (2016-04915).

\bibliographystyle{unsrt}  

\end{document}